\UseRawInputEncoding
\documentclass[twocolumn,aps,amsmath,amssymb,floatfix,superscriptaddress,prl,longbibliography]{revtex4-1}
\usepackage{graphicx}% Include figure files
\usepackage{dcolumn}% Align table columns on decimal point
\usepackage{bm}% bold math
\usepackage{soul}
\usepackage{xspace}
\usepackage{hyperref}
\usepackage{color}
\usepackage[flushleft]{threeparttable}
\def\new{\color{black}}
\def\cb{CrBr$_{3}$\xspace}
\begin{document}

%\setlength{\topmargin}{0in} %%%set margin to be zero to have proper ps printout
%%%when uploaded to arxiv and aps, this has to be commented.
\title{Topological magnon insulator spin excitations in the two-dimensional ferromagnet CrBr$_3$}

\author{Zhengwei~Cai}
\altaffiliation{These authors contributed equally to the work.}
\author{Song Bao}
\altaffiliation{These authors contributed equally to the work.}
\author{Zhao-Long~Gu}
\altaffiliation{These authors contributed equally to the work.}
\author{Yi-Peng~Gao}
\altaffiliation{These authors contributed equally to the work.}
\affiliation{National Laboratory of Solid State Microstructures and Department of Physics, Nanjing University, Nanjing 210093, China}
\author{Zhen~Ma}
\affiliation{National Laboratory of Solid State Microstructures and Department of Physics, Nanjing University, Nanjing 210093, China}
\affiliation{Institute for Advanced Materials, Hubei Normal University, Huangshi 435002, China}
\author{Yanyan~Shangguan}
\author{Wenda~Si}
\affiliation{National Laboratory of Solid State Microstructures and Department of Physics, Nanjing University, Nanjing 210093, China}
\author{Zhao-Yang~Dong}
\affiliation{Department of Applied Physics, Nanjing University of Science and Technology, Nanjing 210094, China}
\author{Wei~Wang}
\affiliation{School of Science, Nanjing University of Posts and Telecommunications, Nanjing 210023, China}
\author{Yizhang~Wu}%yzwu@smail.nju.edu.cn
\author{Dongjing~Lin}%djlin@smail.nju.edu.cn
\affiliation{National Laboratory of Solid State Microstructures and Department of Physics, Nanjing University, Nanjing 210093, China}
\author{Jinghui~Wang}
\author{Kejing~Ran}
\affiliation{School of Physical Science and Technology and ShanghaiTech Laboratory for Topological Physics, ShanghaiTech University, Shanghai 200031, China}
\author{Shichao~Li}
\affiliation{National Laboratory of Solid State Microstructures and Department of Physics, Nanjing University, Nanjing 210093, China}
\author{Devashibhai~Adroja}%devashibhai.adroja@stfc.ac.uk
\affiliation{ISIS Facility, Rutherford Appleton Laboratory, Chilton, Didcot, Oxon, OX11 0QX, United Kingdom}
\affiliation{Highly Correlated Matter Research Group, Physics Department, University of Johannesburg, P.O. Box 524, Auckland Park 2006, South Africa}
\author{Xiaoxiang~Xi}
\author{Shun-Li~Yu}
\email{slyu@nju.edu.cn}
\author{Xiaoshan~Wu}
\author{Jian-Xin~Li}
\email{jxli@nju.edu.cn}
\author{Jinsheng~Wen}
\email{jwen@nju.edu.cn}
\affiliation{National Laboratory of Solid State Microstructures and Department of Physics, Nanjing University, Nanjing 210093, China}
\affiliation{Collaborative Innovation Center of Advanced Microstructures, Nanjing University, Nanjing 210093, China}

\begin{abstract}
Topological magnons are bosonic analogues of topological fermions in electronic systems. They have been studied extensively by theory but rarely realized by experiment. Here, by performing inelastic neutron scattering measurements on single crystals of a two-dimensional ferromagnet CrBr$_3$, which was classified as Dirac magnon semimetal featured by the linear bands crossing at the Dirac points, we fully map out the magnetic excitation spectra, and reveal that there is an apparent gap of $\sim$3.5~meV between the acoustic and optical branches of the magnons at the K point. {\new By collaborative efforts between experiment and theoretical calculations using a five-orbital Hubbard model obtained from first-principles calculations to derive the exchange parameters}, we find that a Hamiltonian with Heisenberg exchange interactions, next-nearest-neighbor Dzyaloshinskii-Moriya (DM) interaction, and single-ion anisotropy is more appropriate to describe the system. Calculations using the model show that the lower and upper magnon bands separated by the gap exhibit Chern numbers of $\pm1$. These results indicate that CrBr$_3$ is a topological magnon insulator, where the nontrivial gap is a result of the DM interaction.
\end{abstract}

\maketitle
Topological electronic systems with nontrivial topological electron band structures and emergent fascinating physical properties have been at the forefront of condensed matter physics in recent decades~\cite{RevModPhys.82.3045,RevModPhys.83.1057,wehling2014dirac,RevModPhys.90.015001}. The concept of band topology can be extended from fermionic to bosonic systems, {\it e.g.}, magnons also~\cite{RevModPhys.88.021004}. In fact, there have been many theoretical proposals for various topological magnon band structures, including topological magnon insulators~\cite{PhysRevB.87.144101,PhysRevB.90.024412,0953-8984-28-38-386001,PhysRevB.97.081106,PhysRevB.97.245111,2020arXiv201106543M}, and magnonic Dirac~\cite{PhysRevB.94.075401,PhysRevLett.119.107205,2399-6528-1-2-025007,PhysRevX.8.011010,sr7_6931,PhysRevLett.119.247202,PhysRevB.97.014433,PhysRevB.99.035160}, and Weyl semimetals~\cite{nc7_12691,PhysRevLett.117.157204,PhysRevB.95.224403,PhysRevB.97.094412}. Similar to fermionic systems, topological magnonic materials also possess exotic topological properties, such as the non-zero Berry curvature responsible for the thermal Hall effect describing the charge-neutral spin excitations carrying heat flowing transversely under external magnetic fields~\cite{PhysRevLett.104.066403,Onose297,PhysRevLett.106.197202,PhysRevB.85.134411,PhysRevB.89.134409,PhysRevLett.115.106603,nc6_6805,Cao_2015,PhysRevB.94.094405,1367-2630-18-10-103039,0953-8984-29-3-03LT01}, and edge or surface state that is topologically protected and robust to perturbations~\cite{PhysRevB.87.144101,PhysRevB.90.024412,PhysRevB.87.174402,PhysRevB.87.174427,PhysRevLett.104.066403,doi:10.7566/JPSJ.85.104707,PhysRevB.97.081106}. Therefore, they hold great potentials in designing next-generation high-efficiency and low-cost magnon spintronics~\cite{RevModPhys.90.015005,chumak2015magnon,PhysRevB.97.081106,PhysRevApplied.9.024029}.  So far, only a handful of topological magnonic states have been realized experimentally, including topological magnon insulator in Cu[1,3-benzenedicarboxylate]~\cite{PhysRevLett.115.147201,PhysRevB.93.214403}, SrCu$_2$(BO$_3$)$_2$~\cite{np13_736} and Ba$_2$CuSi$_2$O$_6$Cl$_2$~\cite{nc10_2096}, and Dirac magnons in CoTiO$_3$~\cite{PhysRevX.10.011062,2020arXiv200703764Y,2020arXiv200704199E} and Cu$_3$TeO$_6$~\cite{Bao2018,np14_1011,PhysRevB.101.214419}. 

\begin{figure}[ht]
\centering
\includegraphics[width=0.98\linewidth]{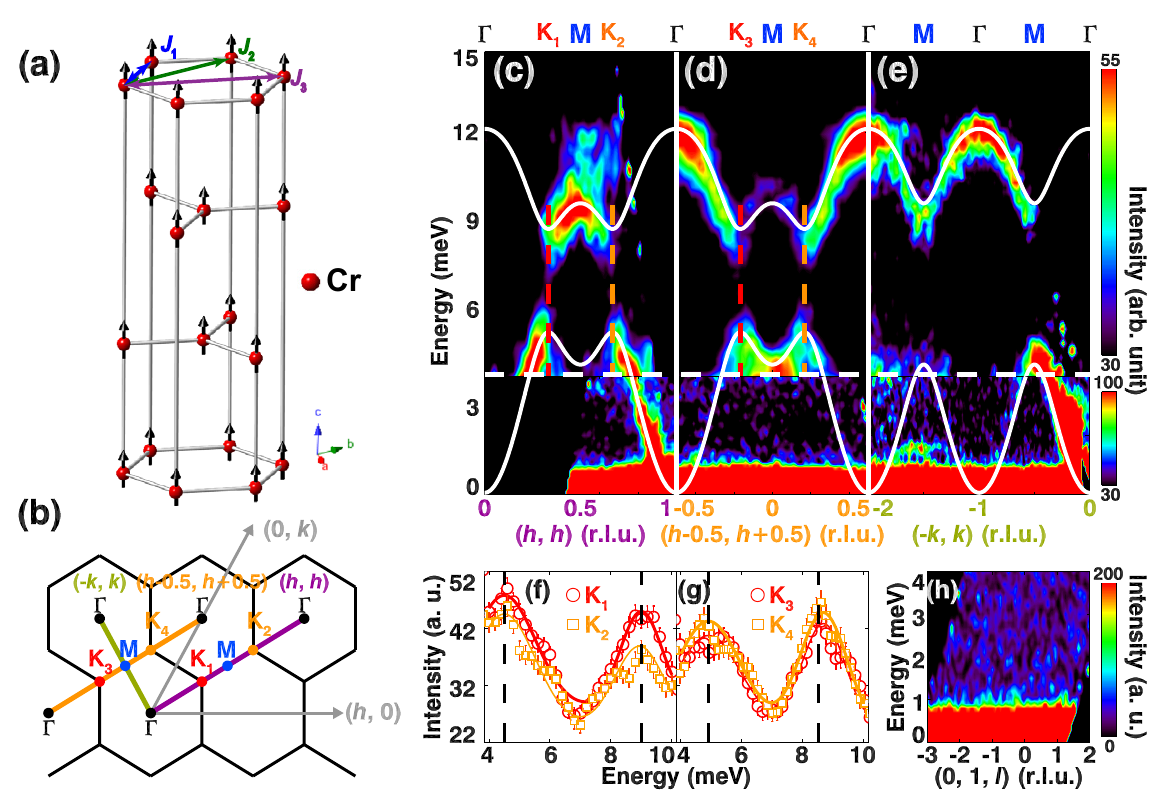}
\caption{ (a) Schematic crystal structure of \cb. Arrows denote the spins. $J_1$, $J_2$ and $J_3$ represent NN, NNN, and NNNN exchange couplings, respectively. (b) Schematic Brillouin zone with high-symmetry points and directions on the ($h$,\,$k$,\,0) plane. (c), (d), and (e) show INS spectra collected at 6~K along ($h$,\,$h$), ($h$-0.5,\,$h$+0.5), and (-$k$,\,$k$) directions, respectively. These directions are illustrated in (b). The high-energy part ($\geq4$~meV, above the horizontal dashed line) was obtained with $E_{\rm i}=30$~meV, while the low-energy part was obtained with $E_{\rm i}=10.7$~meV. The high-energy data were integrated with the orthogonal direction over $[-0.2,\,0.2]$~r.l.u., while the low-energy data were over $[-0.15,\,0.15]$~r.l.u, and $l$ over [-5,5]~r.l.u. for both cases. Solid curves are the calculated dispersions with the DM model, as discussed in the main text. Vertical dashed lines in (c) and (d) illustrate constant ${\bm Q}$ cuts at the K points, and the cut results of (c) and (d) are plotted in (f) and (g), respectively. Solid curves in (f) and (g) are guides to the eye, and the vertical dashed lines denote the peak positions, between which we define to be the gap energy. (h) Magnetic spectrum along the [001] direction obtained with $E_{\rm i}=10.7$~meV. Throughout the paper, errors represent one standard deviation.
\label{fig1}}
\end{figure}

Recently, the honeycomb-structured CrI$_3$, a two-dimensional ferromagnet with a Curie temperature $T_{\rm C}$ of 45~K, not too much reduced from the bulk value of 61~K~\cite{Huang2017,Lado_2017}, has also been suggested to be a topological magnon insulator candidate~\cite{PhysRevX.8.041028,PhysRevB.101.134418}. There are two magnetic atoms per unit cell in this material, and the acoustic and optical branch of the spin waves act as the valence and conduction band respectively in fermionic systems. According to the inelastic neutron scattering (INS) results in Ref.~\onlinecite{PhysRevX.8.041028}, there is a clear gap between the two bands, resulting from the next-nearest-neighbor (NNN) Dzyaloshinskii-Moriya (DM) interaction that leads to the nonzero Berry phase of the bands~\cite{DZYALOSHINSKY1958241,PhysRev.120.91}. Later on, it has been suggested that the bond-dependent Kitaev interaction can also be responsible for the topological magnon band structure~\cite{PhysRevLett.124.017201,PhysRevB.101.134418,PhysRevB.102.115162}. Although the gap opened by either the DM or Kitaev interaction is topological, and thus both results classify CrI$_3$ as a topological magnon insulator~\cite{PhysRevX.8.041028,PhysRevLett.124.017201}, the precise mechanism for the topological band structure is yet to be determined~\cite{PhysRevX.8.041028,PhysRevB.101.134418,PhysRevLett.124.017201,PhysRevB.102.115162,2020arXiv200904475S,2020arXiv201203099J,2020arXiv201213729Z}. Interestingly, another two-dimensional ferromagnet \cb has been labeled as a magnonic Dirac semimetal instead~\cite{PhysRevX.8.011010}, for which the doubly degenerate acoustic and optical branches cross each other and form a Dirac cone at the two-dimensional Brillouin zone corner, K point~\cite{PhysRevB.3.157,PhysRevB.4.2280}, although both its crystal and magnetic structures are similar to those of CrI$_3$~\cite{Wang_2011,cryst7050121}. This provides an excellent opportunity not only to examine the topological state in \cb, but also to pin down the mechanism responsible for the band topology in \cb and CrI$_3$~\cite{PhysRevX.8.041028,PhysRevB.101.134418,PhysRevLett.124.017201,PhysRevB.102.115162,2020arXiv200904475S,2020arXiv201203099J,2020arXiv201213729Z}.

In this Letter, we have carried out comprehensive INS measurements on single crystals of  \cb and fully mapped out the spin-wave excitations in the entire momentum-energy space of interest. We observe a clear gap of $\sim$3.5~meV between the acoustic and optical branches at the K point, where the two modes were expected to cross each other and form Dirac magnons~\cite{PhysRevB.3.157,PhysRevB.4.2280,PhysRevX.8.011010}. {\new By performing theoretical calculations using a five-orbital Hubbard model with tight-binding parameters derived from first-principles calculations, and by careful analysis of the experimental and theoretical data}, we find the material can be more properly described by a Heisenberg Hamiltonian with NNN DM interaction $A$ and single-ion anisotropy constant $D_z$, and the best fit to the experimental spectra yields a parameter set of nearest-neighbor (NN) Heisenberg exchange interaction $J_1=-1.48$~meV, NNN $J_2=-0.08$~meV, next-next-NN (NNNN) $J_3=0.11$~meV, $A=0.22$~meV, and $D_z=-0.02$~meV. Calculations using this model show that the lower acoustic and upper optical magnon bands separated by the gap have a Chern number of $+1$ and $-1$, respectively. These results thus classify CrBr$_3$ as a topological magnon insulator, where the nontrivial gap is opened by the NNN DM interaction.

Single crystals of CrBr$_3$ were grown by the chemical-vapor-transport method. For the INS experiment, we used 40 pieces of single crystals weighed about 1.06~g in total, which were well coaligned using a Laue machine. Time-of-flight INS experiment was performed on a direct geometry chopper spectrometer MERLIN located at ISIS Facility in the United Kingdom. The single crystals were mounted in the ($h$,\,0,\,$l$) plane, with [$\bar{1}$20] aligned along the vertical direction. {\new We used a set of incident energies $E_{\rm i}$ = 30 (primary) and 10.7~meV with a chopper frequency of 350~Hz, which resulted in an energy resolution at the elastic line of 1.16 and 0.35~meV (full width at half maximum), respectively.} The wave vector ${\bm Q}$ was described by ($hkl$) reciprocal lattice unit (r.l.u.) of $(a^{*},\,b^{*},\,c^{*})=(4\pi/\sqrt{3}a,\,4\pi/\sqrt{3}b,\,2\pi/c)$ with $a=b=6.30$~\AA, and $c=18.37$~\AA.

We carried out powder X-ray diffraction measurements at 300~K and the pattern is shown in Supplementary Fig.~1(a)~\cite{sm}, from which we obtain the lattice parameters. \cb crystallizes in a hexagonal structure with the space group $P$3 (\#143) as shown in Fig.~\ref{fig1}(a). In contrast to CrI$_3$ which undergoes a structural phase transition from the high-temperature monoclinic to the low-temperature rhombohedral phase~\cite{cm27_612,PhysRevB.98.104307}, \cb exhibits no structural transition in the Raman spectra in the whole temperature range measured~\cite{sm}. We measured the magnetization and specific heat for \cb and the results are shown in Supplementary Fig.~2~\cite{sm}, from which we find that the system orders ferromagnetically below the $T_{\rm C} = 32$~K, with Cr spins $S=3/2$ aligned approximately along the $c$ axis~\cite{cryst7050121,KUHLOW1975365}.

The spin-wave excitation spectra along three high-symmetry paths ($h$,\,$h$), ($h$-0.5,\,$h$+0.5), and (-$k$,\,$k$) as illustrated in Fig.~\ref{fig1}(b) are plotted in Fig.~\ref{fig1}(c), (d), and (e), respectively. From Fig.~\ref{fig1}(c) and (d), which show spectra along the same direction but at different ${\bm Q}$ points with different magnetic structural factors, we can clearly observe two spin-wave branches. One low-energy branch emerges from the Brillouin zone center $\Gamma$ point, and reaches the maximum at the K point, which we refer to as the ``acoustic" mode. A high-energy branch has a global maximum, local maximum, and minimum at the $\Gamma$, M and K points, respectively. We label it as the ``optical" mode. In Fig.~\ref{fig1}(e), we plot the spectra along ($-k$, $k$), orthogonal to those in Fig.~\ref{fig1}(c) and (d). In Fig.~\ref{fig1}(e), the optical mode with maximum and minimum at the $\Gamma$ and M points, respectively, is very clear. As shown in Fig.~\ref{fig1}(c) and (d), there is a clear gap separating the acoustic and optical branches at the K point. To quantify the gap, we performed constant-$\bm{Q}$ cuts at four K points as illustrated by the dashed lines in Fig.~\ref{fig1}(c) and (d). {\new The results shown in Fig.~\ref{fig1}(f) and (g) clearly exhibit a double-peak profile where the two peaks correspond to the acoustic band top and optical band bottom, with a valley in the middle of the gap.} We fit the cut profiles and take the energy difference between the two peaks as the gap. We average these values at all 24 K points available and obtain a gap energy of $\sim$3.5~meV, smaller than that in CrI$_3$~(Refs.~\onlinecite{PhysRevX.8.041028,PhysRevLett.124.017201}). The gap can also be clearly visualized from Supplementary Fig.~3, where we plot constant-energy contour maps showing the excitation spectra in the ($h$,\,$k$) plane. The observation of a clear gap separating the acoustic and optical bands in \cb is in stark contrast to the expectation that these two bands will cross each other and form a Dirac cone at the K point~\cite{PhysRevB.3.157,PhysRevB.4.2280,PhysRevX.8.011010}. {\new From our results, we also note that there are finite intensities extending from the acoustic and optical branches into the gaped regime due to the resolution effect. We suspect that the intensities might be misinterpreted as resulting from the band crossing in a triple-axis experiment with limited data~\cite{PhysRevB.3.157,PhysRevB.4.2280}.}

In Fig.~\ref{fig1}(h), we plot the spectra along the [001] direction, where we find any meaningful excitations will be overwhelmed by the incoherent elastic scattering. We can thus set an upper limit of 1~meV for the band top of the out-of-plane dispersion. This indicates that the interlayer coupling $J_c$ is much weaker than that in CrI$_3$ where the bandwidth along the [001] direction is $\sim$2~meV~\cite{PhysRevX.8.041028}. Therefore, we neglect $J_c$ in the following discussions for \cb. 

\begin{figure}[ht]
	\centering
	\includegraphics[width=0.98\linewidth]{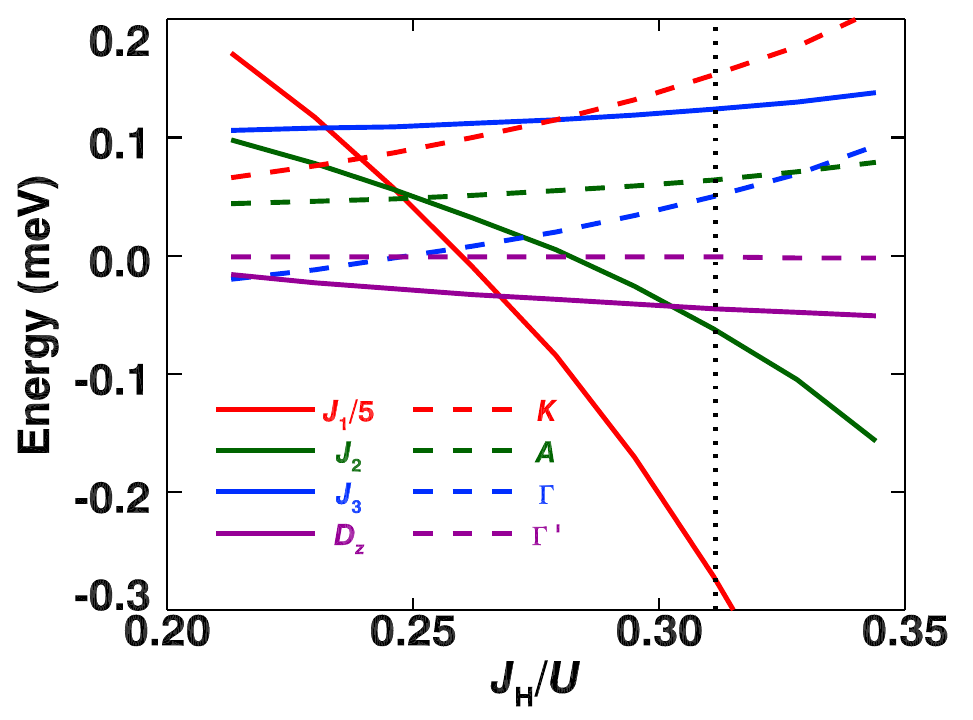}
	\caption{{\new Calculated dependence of the Heisenberg exchange interactions $J_1$, $J_2$, and $J_3$, NN Kitaev interaction $K$, off-diagonal interactions $\Gamma$ and $\Gamma^\prime$, NNN DM interaction $A$, and single-ion anisotropy term $D_z$, as a function of Hund's coupling $J_{\rm H}$ over the Coulomb interaction $U$. The vertical line indicates the material parameters with $U=3.05$~eV and $J_{\rm H}=0.95$~eV. An SOC constant $\lambda=30$~meV was used in the calculations.}
		\label{fig2}}
\end{figure}

\begin{figure*}[ht]
\centering
\includegraphics[width=0.95\linewidth]{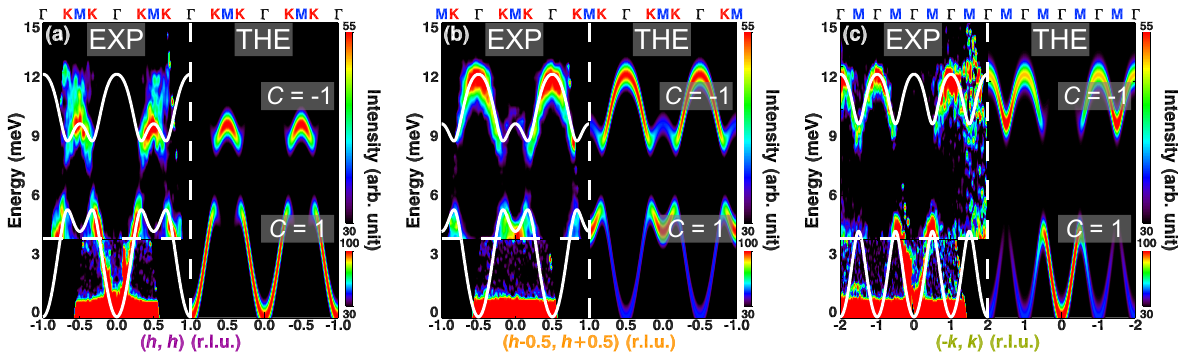}
\caption{Comparison between the experimental (on the left of each panel) and calculated excitation spectra using a Hamiltonian consisted of $J_1$, $J_2$, $J_3$, $A$, and $D_z$ (on the right of each panel). The experimental spectra were obtained under the same conditions as those in Fig.~\ref{fig1}(c)-(e). Calculated dispersions with $J_1=-1.48$~meV, $J_2=-0.08$~meV, $J_3=0.11$~meV, $A=0.22$~meV, and $D_z=-0.02$~meV, are plotted as solid lines on top of the experimental spectra on the left side in each panel. Chern numbers of the acoustic and optical bands are also labeled.
\label{fig3}}
\end{figure*}

Next, we discuss appropriate models to describe the INS excitation spectra. 
First, we can use a Heisenberg Hamiltonian with a DM interaction and single-ion anisotrpy term similar to CrI$_3$~\cite{PhysRevX.8.041028}. In this model, the DM interaction breaks the spatial inversion symmetry and opens the gap at the Dirac point~\cite{DZYALOSHINSKY1958241,PhysRev.120.91,0953-8984-28-38-386001,PhysRevB.97.081106,PhysRevX.8.041028}. In addition, it has been suggested that the anisotropic Kitaev interaction originating from the superexchange processes via strong spin-orbital coupling (SOC) anions can also lift the degeneracy at the Dirac point~\cite{PhysRevLett.124.017201}. {\new To determine whether the DM or Kitaev interaction is responsible for the opening of the Dirac gap~\cite{PhysRevX.8.041028,PhysRevB.101.134418,PhysRevLett.124.017201,PhysRevB.102.115162,2020arXiv200904475S,2020arXiv201203099J,2020arXiv201213729Z}, we use a five-orbital Hubbard model obtained from first-principle calculations to derive the exchange parameters for both CrI$_3$ and CrBr$_3$, as has been done in Refs.~\cite{PhysRevB.96.115103,PhysRevLett.118.107203}, and compare them with the experiment.} The results for both compounds are similar and the details of the calculations are presented in Ref.~\cite{sm}.

The calculated dependence of the exchange parameters on the Hund's coupling $J_{\rm H}$ over the Coulomb interaction $U$ for \cb is shown in Fig.~\ref{fig2}. {\new We used the material parameters with $U=3.05$~eV and $J_{\rm H}=0.95$~eV, which are similar to those obtained from first-principles calculations for CrI$_3$ in Ref.~\onlinecite{PhysRevMaterials.3.031001}. An SOC $\lambda=30$~meV was also included in our work. The corresponding calculated exchange parameters are listed in Table~\ref{tab:para}, where we have a dominant $J_1=-1.36$~meV, an order of magnitude larger than the positive $K$ of 0.15~meV. Considering the small SOC and the large pure spin of 3/2, it is reasonable that the system exhibits a small anisotropic spin interaction $K$. It is worth to note that  Ref.~\cite{2020arXiv200904475S} also found $J_1$ to be orders of magnitude larger than the positive $K$. On the other hand, as shown in Supplementary Fig.~4 and Table~\ref{tab:para}, our best fit to the experimental spectra if using the $JK\Gamma$ model yields a dominant $K=-4.29$~meV while $J_1$ of effectively zero. Similarly, a negative $K$ was also dominating over $J_1$ in the experimental works in CrI$_3$ if the $JK\Gamma$ model was applied~\cite{PhysRevLett.124.017201,PhysRevB.101.134418}. Such an outstanding disagreement in both the order of magnitude of the leading terms and the sign of the $K$ term is very difficult to be reconciled. Therefore, we consider the model with the DM interaction where $J_1$ is most prominent to be more appropriate, which is indeed the case as we show below.}

\begin{table*}[htb]
	\begin{threeparttable}
		\caption{{\new Heisenberg couplings ($J$),  NNN DM interaction ($A$), magnetocrystalline anisotropy constant ($D_z$), NN Kitaev ($K$), and off-diagonal interactions ($\Gamma$) of \cb and CrI$_3$. THE denotes calculated exchange parameters with a five-orbital Hubbard model. EXP-$JAD$ and EXP-$JK\Gamma$ denote the parameters obtained from the best fits to the experimental data using Heisenberg model with DM and Kitaev interactions, respectively.}}
		\label{tab:para}
		\begin{tabular*}{\textwidth}{@{\extracolsep{\fill}}cccccccccccccc}
			\hline \hline
			\begin{minipage}{2cm}\vspace{1mm}Material   \vspace{1mm} \end{minipage} & Method & $J_1$~ (meV) & $J_2$~(meV) & $J_3$~(meV) & $A$~(meV) & $D_z$~(meV) & $K$~(meV)  & $\Gamma$~(meV)\\
			\hline
			\begin{minipage}{2.5cm}\vspace{1mm}~\vspace{1mm} \end{minipage} & THE & -1.36 & -0.06 & 0.12 & 0.07 & -0.04 & 0.15 & 0.05\\
			 \cb & EXP-$JAD$ & -1.48 & -0.08 & 0.11 & 0.22  & -0.02 & ---  & --- \\
			& EXP-$JK\Gamma$ & --- & -0.18 & 0.05 & --- & --- & -4.29  & -0.04 \\
			\hline
			\begin{minipage}{2.5cm}\vspace{1mm}~\vspace{1mm} \end{minipage} & THE~\cite{2020arXiv200904475S} & -5.94 & --- & --- & --- & -0.20 & 0.05 & -0.04 \\
			CrI$_3$ & EXP-$JAD$~\cite{PhysRevB.101.134418} & -2.13 & -0.09 & 0.10 & 0.19  & -0.22 & ---  & --- \\
			& EXP-$JK\Gamma$~\cite{PhysRevLett.124.017201} & -0.2 & --- & --- & --- & --- & -5.2  & -0.07 \\
			\hline\hline
		\end{tabular*}
	\end{threeparttable}
\end{table*}

On the honeycomb lattice, the NN DM interaction cancels out upon space inversion, and only the NNN DM interaction remains finite~\cite{PhysRevX.8.041028}. Thus, the Hamiltonian can be written as,
\begin{equation*}
\emph{H}_\emph{\rm DM}=\sum_{\emph{i}<\emph{j}}\emph{J}_{ij}\bm{S}_\emph{i}\cdot\bm{S}_\emph{j}+\sum_{\langle\langle ij \rangle\rangle}\bm{A}\cdot(
\bm{S}_\emph{i}\times\bm{S}_\emph{j})+\sum_{\emph{j}}\emph{D}_\emph{z}(\emph{S}_\emph{j}^\emph{z})^2.
\end{equation*}
For \cb, a zone center gap for the acoustic mode at the $\Gamma$ point was shown to be 0.08~meV by INS and ferromagnetic resonance measurements in Refs.~\cite{PhysRevB.3.157} and \cite{doi:10.1063/1.1728652}, respectively. This value is in line with the magnetization data plotted in Supplementary Fig.~2, which shows that the easy-axis is roughly along the $c$ axis, and there is a small but finite anisotropy field of 0.35~T between the $c$ axis and the $a$-$b$ plane. Therefore, we use this gap value to constrain $D_z$, which is necessary to stabilize ferromagnetism in thin layers~\cite{Huang2017}.  

Based on the linear spin-wave theory~\cite{0953-8984-27-16-166002}, we calculated the spin-spin correlation function for \cb using this model and compare the results with the experiment in Fig.~\ref{fig3} (see Ref.~\cite{sm} for details on the calculations). In Fig.~\ref{fig3}, the experimental INS data are plotted on the left in each panel, while the calculated spectra with a Hamiltonian consisted of $J$, $A$, and $D_z$ are plotted on the right. The calculated dispersions are also plotted on top of the experimental data as solid curves. Parameters obtained from the best fit using the $JAD$ model are listed in Table~\ref{tab:para}. The small $D_z$ is compatible with the small anisotropy field of 0.35~T shown in Supplementary Fig.~2(b), and the small zone center gap of 0.08~meV in Refs.~\cite{PhysRevB.3.157,doi:10.1063/1.1728652}. {\new By comparing these values with the calculated results in Table~\ref{tab:para}, we believe the agreement between theory and experiment is remarkable, although the fitted value of the sub-leading term $A$ is still three times larger than the calculated one, which can be further improved in principle as we discuss in Ref.~\cite{sm}. It is sufficient to justify the $JAD$ model rather than the $JK\Gamma$ model as the appropriate model to describe the material by evaluating the order of magnitude in the leading interactions and the sign of the $K$ term.} It is clear that the model can reproduce the magnetic excitation spectra very well, especially with a clear gap at the K point consistent with the experimental data. As discussed in Ref.~\cite{sm}, we have calculated the Chern numbers~\cite{PhysRevB.87.144101,PhysRevB.97.081106,PhysRevB.52.4223,PhysRevB.87.174402,PhysRevB.87.174427,PhysRevB.99.054409} of the magnon spectra obtained using this model. The results show that both the acoustic and optical bands are topologically nontrivial, having a Chern number +1 and -1, respectively, which are labeled in Fig.~\ref{fig3}.

These results demonstrate that \cb is a new topological magnon insulator, similar to CrI$_3$~\cite{PhysRevX.8.041028}, but in contrast to the theoretical prediction of it being a Dirac magnon~\cite{PhysRevX.8.011010}. The overall energy scale of \cb is reduced from that of CrI$_3$~\cite{PhysRevX.8.041028,PhysRevB.101.134418}, consistent with the smaller $T_{\rm C}$~\cite{Huang2017,Lado_2017}. We can estimate the relative strength of the SOC in CrI$_3$ and \cb by analyzing the ratio of the gap at the K point over the bandwidth. For a more accurate estimation, we use the bandwidth of the optical branch, which can be determined more precisely. For CrI$_3$ and \cb, we have their ratios to be 1.20 and 1.02, respectively. This suggests that the former has a larger SOC due to the heavier anion mass. {\new Overall, as summarized in Table~\ref{tab:para}, our theoretical calculations and those in Ref.~\cite{2020arXiv200904475S}, the exchange parameters extracted from our and others' experiments~\cite{PhysRevX.8.041028,PhysRevB.101.134418,PhysRevLett.124.017201}, the strength of the SOC, and the large and pure spin of 3/2 in Cr$^{3+}$, all indicate that $J_1$ is the leading term and the DM interaction is responsible for opening the gap and making the material a topological magnon insulator. There are further works that can testify this conclusion, such as to apply an in-plane magnetic field and explore the evolution of the gap at the K point, and to measure the thermal Hall and magnon Nernst effects~\cite{2020arXiv201213729Z}.}

In summary, our comprehensive INS excitation spectra clearly unveil a gap of $\sim$3.5~meV between the acoustic and optical branches of the spin-wave excitations at the supposedly Dirac points in a two-dimensional ferromagnet \cb. {\new By carefully analyzing the experimental and theoretical data, we find that a Heisenberg Hamiltonian with a DM interaction and single-ion anisotropy term is more appropriate to describe the system. The DM interaction is responsible for opening the topologically nontrivial gap,  separating the two bands which exhibit nonzero Chern numbers. These results show that \cb is a topological magnon insulator, which should exhibit topological thermal Hall and magnon Nernst effects.}

We would like to thank Hao Su and Prof. Yanfeng Guo at ShanghaiTech University for allowing us to use their Laue machine for sample alignment. The work was supported by the National Natural Science Foundation of China with Grants No.~11822405, 11674157, 12074174, 12074175, 11674158, 11774152, 11904170, 11774151, 11874200, 12004249, and 12004251, National Key Projects for Research and Development of China (Grant No.~2016YFA0300401), Natural Science Foundation of Jiangsu Province with Grants No.~BK20180006 and BK20190436, Shanghai Sailing Program with Grant No. 20YF1430600, Fundamental Research Funds for the Central Universities with Grants No.~020414380117 and 020414380147, and the Office of International Cooperation and Exchanges of Nanjing University. The experiment at ISIS Facility of Rutherford Appleton Laboratory was performed under a user program (Proposal No.~RB1820157), with financial support by the Newton Fund for China. Z.M. thanks Beijing National Laboratory for Condensed Matter Physics for funding support.

\end{document}